\documentstyle[twoside,fleqn,espcrc2,epsf]{article}

\newcommand{\AmS}{{\protect\the\textfont2
  A\kern-.1667em\lower.5ex\hbox{M}\kern-.125emS}}
\title{Towards the Deconfinement Phase Transition in 
Hot Gauge Theories 
\thanks{Work supported by Bundesministerium f\"ur Wissenschaft,
und Transport of Austria}
        \thanks{presented by O. Borisenko}}

\author{O. Borisenko, M. Faber\address{Institut f\"ur Kernphysik, 
Technische Universit\"at Wien, A-1040 Vienna, Austria}%
      \  and G. Zinovjev\address{Institute for Theoretical Physics, 
National Academy of Sciences of Ukraine, 
Kiev 252143, Ukraine}}

\begin{document}

\begin{abstract}
The phase structure of hot gauge theories with dynamical matter fields
is reexamined in the canonical ensemble with respect to triality. 
We discuss properties of chromoelectric
and chromomagnetic sectors of the theory and show whereas electric
charges carrying a unit of $Z(N_c)$ charge are screened at high temperatures
via dynamical matter loops, this is not the case for the $Z(N_c)$ magnetic
flux. An order parameter is constructed to probe the realization
of {\it local} $Z(N_c)$ symmetry in the magnetic sector. We argue this
order parameter may be used to detect the deconfinement phase transition
which is defined in terms of the screening mechanism.
\end{abstract}

\maketitle

Here we continue an investigation of the phase structure of hot
gauge theories in the canonical ensemble with respect to triality
\cite{trl}-\cite{preprint}.
Usually, the deconfinement phase transition 
is associated with the appearence of nonzero triality states
in hot phase \cite{md,mo}. 
In particular $Z(N_c)$ gauge theories with Higgs fields 
have been precisely analyzed \cite{mo}.
Since it was known that at zero temperature this system
has two phases, a confining/screening phase and a deconfining one
\cite{mm} separated by a critical line and since such
a critical line was not found it has been
concluded that at finite $T$ the critical behaviour 
may not be present at all. 

Our motivation for what follows is

I) it is not obvious a priori that ``free triality states''
indeed exist in deconfined phase and quite possible 
there might be a phase 
transition unrelated to the triality liberation 
but rather to {\it different screening mechanisms of triality};

II) the previous emphasis has been put on
the realization of global $Z(N_c)$ symmetry.
Localizing $Z(N_c)$ looks promising \cite{kraus2}.

We are dealing with the canonical ensemble (CE) 
introduced in \cite{trl,versus}. This ensemble reveals the
following properties \cite{preprint}: 
		
1) In the low temperature phase every state has zero triality.
In the deconfined phase the whole system possesses zero triality.
Since all $Z(N_c)$ noninvariant variables are projected out,
the Polyakov loop (PL) itself has only little meaning in the CE and
a single quark does not appear in the spectrum;

2) Metastable minima with unphysical properties are absent, all $Z(N_c)$
phases are degenerate;

3) Chiral symmetry is restored in all $Z(N_c)$ phases and at 
the same temperature \cite{preprint}.

Generalizing the $A$ operator introduced in \cite{kraus2} 
to finite $T$ theory, correlation function of PLs should be considered
instead of Wilson loop (WL) as
\begin{equation}
A_t(\Sigma , R) = \frac{<L_0 L_R>_F}{<L_0 L_R>_0} \ ,
\label{ZNOP1}
\end{equation}
\noindent
where the numerator is calculated over a frustrated ensemble
defined by the partition function in CE
\begin{equation}
Z_F = \frac{1}{N_c} \sum_{k=1}^{N_c} \int \prod_l dU_l \prod_{x,i}
d\bar{\Psi}_x^i d\Psi_x^i e^{-S_F - S_q(k)}
\label{Fpf}
\end{equation}
\noindent
$S_q$ is a quark action, where
$U_0\rightarrow \exp [\frac{2\pi k}{N_c}]U_0$ for a time slice 
\cite{trl} and ``frustrated'' action is obtained from the Wilson action 
$S_W$ as $S_W\to S_F=S_W(ZU_p)$.  Singular transformations $Z$
are defined on a closed dual surface $\Sigma$ \cite{preprint}.
A similar $F$-ensemble may be constructed for spatial WL
and the $A_s$ operator which is the same as in zero temperature
theory introduced. 
The $A_t$ and $A_s$ can be used to measure screening effects of 
dynamical fields and screening effects of pure gluonic interaction 
estimating their competition.

If $W(C)$ is WL in $Z(N_c)$ pure gauge theory one has 
in the weak coupling regime
\begin{equation}
<W(C)> \sim \exp (-\gamma_{gl} P_C),
\label{pergauge}
\end{equation}
\noindent
$P_C$ is a perimeter of loop $C$.
Matter fields enforce WL
to decay with perimeter law at {\it any} coupling
\begin{equation}
<W(C)> \sim \exp (-\gamma_{dyn} P_C).
\label{perhiggs}
\end{equation}
\noindent
We refer to $\gamma_{dyn}$ as coming from the dynamical
screening. Keeping the system in a coupling region
where the pure gauge interaction leads to an
area law one has 
$$
<W(C)> \sim K_1\exp (-\gamma_{dyn} P_C) + K_2\exp (-\alpha_{gl} S).
$$
\noindent
$\alpha_{gl}$ is the string tension of pure gauge theory.
In the  $F$ ensemble $K_2 \to - K_2$. It gives $A=1$ at
$C\to\infty$ and signals that dynamical
screening dominates the system. There is a critical 
point in a pure gauge system above which 
$$
<W(C)> \sim K\exp (-\gamma_{dyn} P_C)+K_4\exp (-\gamma_{gl} P_C).
$$
\noindent
In the $F$ ensemble one has to change the sign of $K_4$.
There is a competition between the kinetic and
dynamical screening. One gets
\begin{equation}
A(\Sigma , C) = \left\{
\begin{array}{c}
1 , \gamma_{dyn} < \gamma_{gl} , \\
-1, \gamma_{dyn} > \gamma_{gl} .
\end{array}
\right\} \ .
\label{Atodeconf}
\end{equation}
In the lower regime the kinetic screening gets stronger and
$Z(N_c)$ charge can be detected. This is
an {\it inherent} feature of the deconfined phase.
In fact, (\ref{Atodeconf}) predicts exact equation for the critical 
line in the theory $\gamma_{dyn}(\alpha) = \gamma_{gl}(g^2)$
with $g^2$ and $\alpha$ the gauge and Higgs couplings,
respectively. 

We argue, to reveal the critical 
behaviour one has to analyze screening mechanisms of triality
in different {\it gauge coupling} intervals.
We {\it define} a deconfinement phase of the theory with
dynamical matter fields as {\it a weak coupling phase where the  
screening due to gluon interactions is stronger than the 
dynamical one}. 

At finite $T$ the spatial WL
behaves as at $T=0$ and $A_s$ 
should be a proper order parameter, too.
A nontrivial value of $A_s$ implies that a unit of $Z(N_c)$ flux
is unscreened dynamically and detectable at long range.

The behaviour of $A_t$ differs.
In strong coupling regime the correlation function of
PLs for pure gauge sector decays exponentially. 
The fermionic sector generates terms screening
heavy quarks and leads to a constant value of the correlator
even at spatial infinity. This implies $A_t=1$.
In weak coupling region the pure gauge sector also
gives a finite value for the correlator at spatial infinity
leading to a competition with the dynamical
screening. Hence, the direct use of $A_t$ 
as an indicator of a phase transition is  
impossible because both contributions are finite in the
$R\to\infty$ limit. One may argue that $A_t=1$, 
the stable interfaces of pure gauge system become {\it unstable}
in the presence of dynamical matter.

Applying above idea we examine 
the model of $Z(2)$ gauge spins coupled to the
Higgs fields at finite $T$. 
The canonical partition function of the $Z(2)$ gauge model
is given by the path integral
\begin{equation}
Z = \frac{1}{2} \sum_{k= \pm 1} \sum_{s_l=\pm 1}
 \sum_{z_x=\pm 1} \; e^{S_W + S_H} \ ,
\label{z2higgs}
\end{equation}
\noindent
\begin{equation}
S_W = \sum_{p_0} \lambda_0 S_{p_0} + \sum_{p_n} \lambda_n S_{p_n} \ ,
\label{z2g}
\end{equation}
\noindent
\begin{equation}
S_H = S_H^{sp}+S_H^t = \sum_{x,\mu}h_{\mu} z_xs_{\mu}(x)z_{x+\mu}.
\label{higgsaction}
\end{equation}
\noindent
Both fields obey periodicity conditions. To calculate the operator
$A_t$ we shift the surface $\Sigma$ 
to the Higgs part of the action getting
\begin{equation}
A_t(\Sigma , R) = - \ \frac{<L_0 L_R>_F}{<L_0 L_R>_0} \ , \
L_0 = \prod_{t=1}^{N_t}s(0,t).
\label{Athiggs}
\end{equation}
\noindent
Putting $\Omega$ as a volume enclosed by $\Sigma$ we introduce
\begin{equation}
h_0\rightarrow h_0(x)=\left\{
\begin{array}{c}
h_0(x\notin \Omega ),  \\
-h_0(x\in \Omega )
\end{array}
\right\} \ .
\label{heff}
\end{equation}
\noindent

Then if $\lambda_0 \ , \lambda_n << 1$,
using strong coupling expansion
we get (up to the second order)
$$
<L_0L_R> = \prod_{t=1}^{N_t}\tanh h_0(0,t)\tanh h_0(R,t) 
$$
\begin{equation}
[1 + 2DN_t \tanh \lambda_0 \tanh^2h_n (1-\tanh^2h_0) ],
\label{crPL1}
\end{equation}
\noindent
where $D$ is the space dimension. Since the linking number of PL
in the origin and the surface $\Sigma$ is 1, 
it gives $A_t = 1 + o(\lambda^2)$.
The expression in the square brackets  
is an even function of $h_0(t)$ up to the 
$(\tanh \lambda_0)^{L_{\Sigma}}$ order, where $L_{\Sigma}$ is a linear
size of domain enclosed by $\Sigma$. The corresponding
plaquettes will change signs only on the boundary. Thus,
\begin{equation}
A_t = 1 - N_{\Sigma}C_1(\tanh \lambda_0)^{L_{\Sigma}},
\label{Atsc}
\end{equation}
\noindent
where $N_{\Sigma}$ is the number of frustrated plaquettes.
It leads at $\Sigma\to\infty$ to the expected result $A_t=1$.

Now if $\lambda_0 \ , \lambda_n >> 1$, it
seems the fermionic contribution
is suppressed as $h_0^{2N_t}$ and might
be dropped relatively to the kinetic screening but it misleads.
To skip this contribution, one should isolate 
it expanding the correlation function in small $h_0$ what is known
to be divergent \cite{mo}.
Thus, there is no direct way to separate the Debye screening from 
fermion screening in electric sector. 
In the $F$ ensemble there is a competition between the vacuum state $I$
with all the spins up or down depending on the triality sector
and the state $II$ when all the time-like spins are flipped inside 
$\Omega (\Sigma)$ relatively to links outside and the difference
of classical actions gives
\begin{equation}
S^I - S^{II} = 2\lambda_0 N_{\Sigma} - 2h_0 \Omega (\Sigma).
\label{Iact}
\end{equation}
\noindent
On a finite lattice there always exists a large $\lambda_0$
when the surface term wins and $A_t = -1$. 
When $\Sigma \to\infty$ 
this state can be a metastable state only as the volume 
term supresses the surface term in this limit. 
Hence, this is the state from $S^{II}$ which 
dominates the thermodynamic limit and leads to $A_t = 1$.  

Turning to $A_s$ we have as above
\begin{equation}
A_s(\Sigma_s , C) = - \ \frac{<W_s(C)>_F}{<W_s(C)>_0} \ ,
\label{Ashiggs}
\end{equation}
\noindent
where $W_s(C)$ is the space-like WL  
defined in (\ref{z2higgs}), $\Sigma_s$ is a two 
dimensional surface on a dual lattice, $\Omega_s$ is the corresponding 
volume. The temporal part of the Higgs action is not affected by $Z(2)$
singular gauge transformations but for the spatial part
we get in the $F$ ensemble
\begin{equation}
S_H^{sp} = \sum_{x,n}h_n(x) z_x s_n(x) z_{x+n} \ ,
\label{higgsspF}
\end{equation}
\noindent
where $h_n(x)$ is defined similarly to (\ref{heff}). 
If $\lambda_n << 1$ one gets as in previous case 
\begin{equation}
A_s = 1 - N_{\Sigma_s}C_2(\tanh \lambda )^{L_{\Sigma_s}},
\label{ASsc}
\end{equation}
\noindent
which gives $A_s=1$ in the $\Sigma_s\to\infty$ limit.
But if $\lambda_n >> 1$,
the gauge part gives the following contribution to the WL
\begin{equation}
<W(C)> \propto \exp \left[ - 2P_C (e^{-2\lambda})^6 \right].
\label{z2scor}
\end{equation}
\noindent
The fermionic screening is suppressed as $h_n^{P_C}$, i.e.
\begin{equation}
<W(C)> \propto (\tanh h_n)^{P_C}.
\label{z2sferm}
\end{equation}
\noindent
It allows us to expand in small $h_n$ 
since there are no loops going around the lattice in space direction which 
could destroy the convergence. It is straightforward to calculate, 
e.g. $<F(\Sigma_s)>$ in leading order of small $h$
\begin{equation}
<F(\Sigma_s)> = \exp \left[ -\delta N_{\Sigma_s} + O(h^6) \right],
\label{z2F}
\end{equation}
\noindent
where $\delta \approx 2h^4 \tanh \lambda$.
If we shift surface $\Sigma_s$ back to the pure gauge action, 
the dominant contribution in the $F$ ensemble comes from configurations
of gauge fields $s_n(x)$ flipped in the volume $\Omega (\Sigma_s)$
relatively to $s_n(x)$ outside of $\Omega (\Sigma_s)$. 
The WL changes sign in the $F$ ensemble and we find
\begin{equation}
A_s(\Sigma_s , C) = -I.
\label{z2ZNOP}
\end{equation}
\noindent
The critical line in the main order is determined from
$e^{-2(e^{-2\lambda^c})^6} = \tanh h_n^c$.

Thus, $A_t$ probes directly domain walls in finite $T$ theory
and shows whether interfaces are stable. 
$A_s=-1$ indicates a deconfinement phase transition 
to a phase where the kinetic screening dominates
the dynamical one.


\begin{thebibliography}{99}

\bibitem{trl} M.~Faber, O.A.~Borisenko, G.M.~Zinovjev,
Nucl.Phys. B444 (1995) 563.
\bibitem{versus}  M.~Oleszczuk and J.~Polonyi, preprint TPR 92-34, 1992.
\bibitem{preprint} M.~Faber, O.A.~Borisenko, G.M.~Zinovjev,  
Nucl.Phys. B (Proc.Suppl.) 42 (1995) 484; 53 (1997) 462.
Mod.Phys.Lett. A12 (1997) 949.
\bibitem{md} C.~DeTar, L.~McLerran, Phys.Lett. B119 (1982) 171.
\bibitem{mo} H.~Meyer-Ortmanns, Nucl.Phys. B230 (1984) 31. 
\bibitem{mm} R.~Marra, S.~Miracle Sole, Comm.Math.Phys. 67 (1978) 233.
\bibitem{kraus2} J.~Preskill, L.M.~Krauss, Nucl.Phys. B341 (1990) 50.

\end{thebibliography}
\end{document}